\documentclass[aps,prl,reprint,groupedaddress]{revtex4-2}

\usepackage{amsmath}
\usepackage{graphicx} 
\usepackage{siunitx}
\usepackage{soul}
\usepackage{color}

\begin{document}


\title{Scaling law for cracking in shrinkable, granular packings}


\author{H. Jeremy Cho}
\thanks{Current address: \\Department of Mechanical Engineering, University of Nevada, Las Vegas, Las Vegas, NV 89154}

\author{Sujit S. Datta}
\thanks{To whom correspondence should be addressed:\\ssdatta@princeton.edu}
\affiliation{Department of Chemical and Biological Engineering, Princeton University, Princeton, NJ 08544}


\date{\today}

\begin{abstract}
Hydrated granular packings often crack into discrete clusters of grains when dried. Despite its ubiquity, accurate prediction of cracking remains elusive. Here, we elucidate the previously overlooked role of \emph{individual} grain shrinkage---a feature common to many materials---in determining crack patterning using both experiments and simulations. By extending the classical Griffith crack theory, we obtain a scaling law that quantifies how cluster size depends on the interplay between grain shrinkage, stiffness, and size---applicable to a diverse array of shrinkable, granular packings.

\end{abstract}


\maketitle


Hydrated packings of grains often crack when they dry; familiar examples of this phenomenon are cracks in mud and paint \cite{goehring2015desiccation}. Cracking is often undesirable, damaging structures built on clay-rich soil, causing waste leakage from subsurface clay barriers \cite{doi:10.2113/gselements.5.2.105}, altering the texture of foods \cite{kidmosecarrot99}, disrupting biological tissues \cite{ARROYO20171}, and limiting the performance of coatings \cite{0034-4885-76-4-046603}. In other cases, cracking is desirable, but must be controlled; emerging examples include micro/nano-patterning \cite{Kim:2016bh} and transport in fuel cells \cite{Kim:2016gr}. However, despite the ubiquity and practical importance of this phenomenon, accurate prediction of cracking remains elusive.

Cracks often form two-dimensional (2D) polygonal patterns that separate discrete clusters of grains \mbox{\cite{doi:10.1063/PT.3.2584}}. The size of these clusters impacts macroscopic appearance, texture, transport, and mechanics; thus, it is important to identify the factors that affect cluster size. The characteristic cluster size has been found to vary widely depending on drying dynamics, packing size, friction with a substrate, capillarity, and grain-scale factors such as grain stiffness and size \cite{B922206E,doi:10.1021/la061251+,C3SM52528G,PhysRevLett.74.2981,doi:10.1021/la049020v,Groisman_1994,Komatsu_1997,doi:10.1021/la048298k,PhysRevLett.98.218302,doi:10.1098/rsta.2012.0352,Scherer2015,C7SM00985B,PhysRevE.99.012802}. Recent work has taken a first step towards predicting these dependencies by applying classical Griffith crack theory, in which  cracking is governed by a balance of strain and surface energies \cite{doi:10.1098/rsta.1921.0006,Flores:2017js}, to granular packings \cite{Groisman_1994,Komatsu_1997,doi:10.1021/la048298k,PhysRevLett.98.218302,doi:10.1098/rsta.2012.0352,Scherer2015,C7SM00985B,PhysRevE.99.012802}. However, while this approach can describe how the characteristic cluster size scales with packing thickness \cite{PhysRevE.99.012802}, quantitative prediction of cluster size remains elusive for many materials. 

Here, we reveal a key, previously overlooked, granular property that also governs cracking: the ability of \emph{individual} grains to shrink as they dry. Grain shrinkage is common to many materials including clays, soils, coatings, biological tissues, and foods, for which the stresses resulting from shrinkage often dominate over externally applied stresses \cite{Creton_2016,Tanner2003}. Nevertheless, while some models account for macroscopic shrinkage during drying by considering grain densification and deformation \cite{Groisman_1994,Komatsu_1997,doi:10.1021/la048298k,PhysRevLett.98.218302,doi:10.1098/rsta.2012.0352,Scherer2015,C7SM00985B}, only recently has individual grain shrinkage been recognized as a factor that can strongly affect cracking---often in unexpected ways \cite{Cho2019cracks1}. Unfortunately, how grain shrinkage influences the final cluster size after drying is unknown; no current theory of cracking incorporates this behavior. 

In this Letter, we use experiments and discrete-element (DEM) simulations to study the influence of grain shrinkage on crack patterning. The sizes of clusters formed by cracking depend on an interplay between grain stiffness and size, the overall packing size, and capillarity---all of which evolve with grain shrinkage--as well as substrate friction. By explicitly incorporating grain shrinkage into classical crack theory, we show that the cluster sizes can be predicted by balancing the mechanical energy required to break capillary bridges at the boundary of a cluster and the strain energy resulting from grain shrinkage and substrate friction. Our work thus provides a new scaling law that can describe crack patterning in a wide range of soft materials with shrinkable components.


\begin{figure*}
\includegraphics[width=0.85\linewidth]{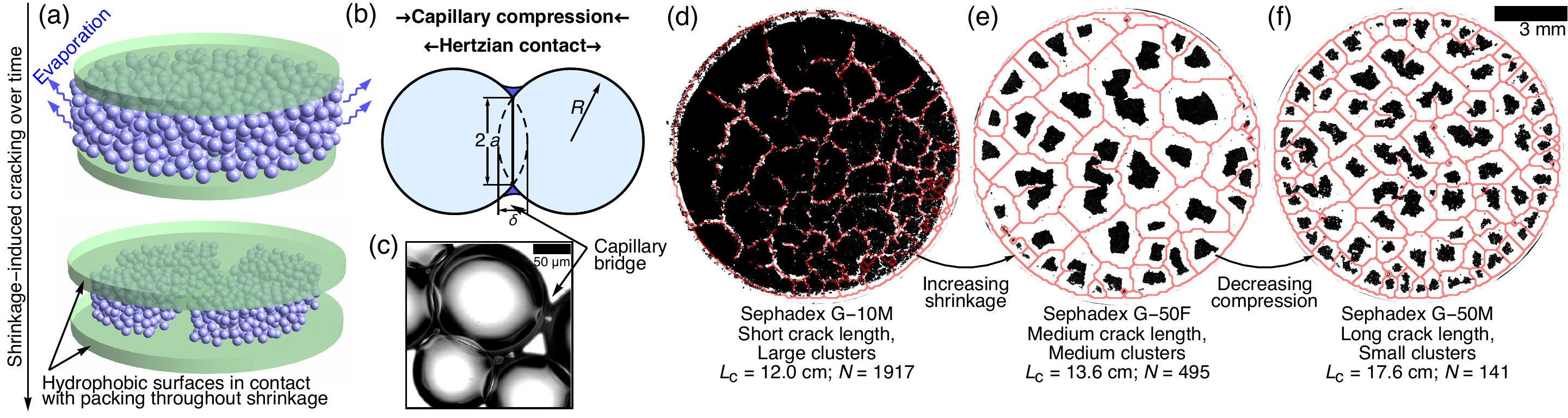}
\caption{Experiments highlight the impact of bead-scale properties on cracking. (a) Schematic of a packing of shrinkable beads confined between two substrata. As the packing dries, the individual beads shrink, leading to the formation of a 2D crack network. (b) The beads are in a pendular state with capillary bridges that compress adjacent beads together (c) as observed via brightfield microscopy. Cracks form when bridges break due to bead shrinkage and friction with the substrate. (d) A packing of small, low shrinkability beads has minimal cracking. (e) A packing of larger, high shrinkability beads has more cracking. (f) A packing of even larger, high shrinkability beads with less compression has even more cracking. We use microscopy to directly measure $\Phi=\num{2.0\pm0.1},~\num{8.5\pm1.1},~\num{8.5\pm1.1}$; $\delta_\text{wet}=\num{1\pm1},~\num{21\pm6},~\SI{15\pm8}{\micro\meter}$; and $R_\text{wet}=\num{34\pm9},~\num{59\pm14},~\SI{86\pm18}{\micro\meter}$ for (d--f), respectively. Parentheses indicate uncertainty.}
\label{fig:fig1}
\end{figure*}

To experimentally test the role of individual grain shrinkage, we study granular packings of shrinkable, non-Brownian, cross-linked hydrogel beads with tunable bead shrinkability, stiffness, and size. Each packing is confined between two smooth, hydrophobic, parallel substrata in a disk geometry with initial packing radius $\mathcal{R}$, chosen to be sufficiently small to ensure uniform bead shrinkage during drying \cite{supp}. The top surface is free to track the top of the packing as it dries, resulting in an unloaded configuration (Figure~\ref{fig:fig1}a). Moreover, to isolate the influence of bead shrinkage, we prepare each packing in an initial pendular configuration \cite{doi:10.1080/00018730600626065}, with the hydrated beads held together by capillary bridges as previously described \cite{Cho2019cracks1}. The capillary bridges initially compress the individual beads by a distance $\delta_\text{wet}$ as shown in Fig.~\ref{fig:fig1}b--c \cite{Moller:2007bp,Butt:2010gs,Herminghaus:2005ei}. As the packings continue to dry, the beads shrink in volume by a factor $\Phi \equiv R_\text{wet}^{3}/R_\text{dry}^{3}$ where $R$ is the radius of an individual bead and the $\text{wet}$ and $\text{dry}$ subscripts refer to the initial hydrated and final dehydrated states, respectively. Often known as the degree of swelling, we term $\Phi$ the bead \emph{shrinkability} to emphasize the  shrinkage during the drying process. The combination of bead shrinkage and substrate friction then causes beads to separate, breaking the capillary bridges between some of them, leading to the formation of cracks \mbox{\cite{doi:10.1021/la9903090}}; these bridges can thus be weaker than the inter-atomic/molecular bonds in continuum solids. The friction results from capillary adhesion to the substrate; thus, it exists in the absence of any load unlike classical friction \mbox{\cite{Groisman_1994}}. In uncracked regions, the beads remain held together by capillary bridges, even at the final dry state, likely due to non-zero humidity in the ambient.

Cracks ultimately form a 2D network that separates discrete, polygonal clusters of close-packed grains (Fig.~\ref{fig:fig1}). The cluster geometries are uniformly extruded along the packing height. We thus apply a morphological thinning algorithm to binarized 2D projections to identify the crack network, shown by the red lines in Fig.~\ref{fig:fig1}d--f. Measuring the total length of this network yields the total crack length $L_\text{c}$, and summing the individual pixels circumscribed by the network yields the effective cluster area at the initial hydrated state, $A$. These clusters tessellate the entire extent of the packing. Assuming uniform clusters then yields a characteristic cluster area, defined as the number of beads making up the hydrated cluster area $A$, $N \sim \mathcal{R}^4/L_\text{c}^{2}R_\text{wet}^{2}$ \cite{supp}.


First, we test packings confined by the same substrata---thus, having similar substrate friction---and having the same height. Classical crack theory predicts that cluster size varies with these two factors; however, previous work do not consider the influence of grain shrinkability $\Phi$. To test the influence of this critical parameter, as well as the other grain-scale parameters $\delta_\text{wet}$ and $R_\text{wet}$, we test four different classes of beads made of the same material: Sephadex G-10M, G-25M, G-50M, and G-50F. These have varying hydrated mesh sizes, denoted by the number, resulting in shrinkability $\Phi$ increasing from 2.0 to 8.5 and capillary compression $\delta_\text{wet}$ increasing from \num{1} to \SI{21}{\micro\meter}, respectively. They also have varying hydrated bead sizes $R_\text{wet}$, denoted by the letter M or F, ranging between \num{34} and \SI{86}{\micro\meter}. 

We observe striking differences in cracking for the different packings. The total crack length $L_\text{c}$ increases, and the equivalent cluster area $N$ decreases, with increasing $\Phi$ and $R_\text{wet}$; compare Fig.~\ref{fig:fig1}d,e. The crack length also increases, and the cluster area decreases, with decreasing $\delta_\text{wet}$ and increasing $R_\text{wet}$; compare Fig.~\ref{fig:fig1}e,f. Thus, different packings with the same height and substrate friction can have vastly differing cluster sizes depending on grain shrinkage, stiffness, and size.

\begin{figure*}
\includegraphics[width=0.89\linewidth]{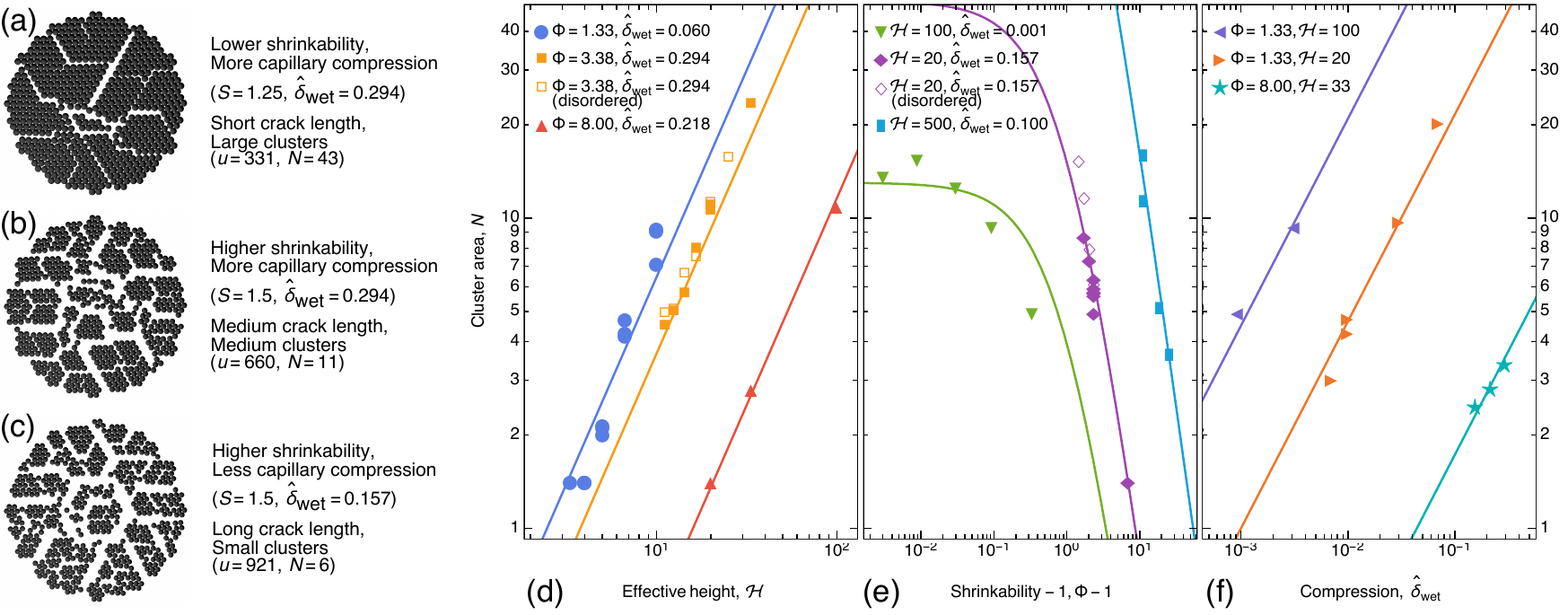}
\caption{DEM simulations isolate the influence of bead shrinkage, stiffness, and size on cracking. (a) A simulated packing of low shrinkability beads has minimal cracking. (b) A simulated packing of high shrinkability beads has a larger degree of cracking. (c) A simulated packing of high shrinkability beads with less capillary compression has an even larger degree of cracking. In (a--c), we set $M = 613$ and hold $F_\text{fr} / \left(2\pi\gamma R_\text{wet}\right) = 0.05$ fixed to mimic the experiments. In (d--f), we quantify the equivalent cluster area, $N$, while independently varying the (d) effective height $\mathcal{H}$, (e) shrinkability $\Phi$, and (f) bead compression $\hat{\delta}_\text{wet}$. Colored lines are curves fit to the data indicating that $N$ (d) increases with height as $\propto \mathcal{H}^{4/3}$, (e) decreases with the shrinkability as $\propto \Phi^{-\left(2m/3+2/9\right)}$ with $m=9/4$, and (f) increases with bead compression as $\propto \hat{\delta}_\text{wet}^{2/3}$. We do not observe a considerable difference in $N$ upon varying packing size, $M$, from 91 to 823 or using a disordered packing configuration (open symbols).}
\label{fig:demresults}
\end{figure*}


To further investigate the role of these factors on cracking, we next perform DEM simulations of 2D packings of shrinkable beads. We specify $\Phi$ in each simulation. The beads are interconnected by capillary bridge bonds in the pendular configuration, and shrink from the initial hydrated state ($R=R_\text{wet}$) to the final dehydrated state ($R=R_\text{dry}$) as drying progresses; thus, these bonds are unlike bonds in continuum solids, which do not change during drying. During drying, the inter-bead Hertzian contact force resisting compression $\sim K R^{1/2} \delta^{3/2}$, where $K$ and $\delta$ are the shrinkage-dependent bead bulk modulus and compression, respectively, as supported by previous work \cite{supp,C003344H,doi:10.1021/ma50002a038,doi:10.1002/app.1993.070500619,Andrei1998}. Conversely, the capillary force compressing the beads together is $2\pi \gamma a$, where $\gamma$ is the surface tension of water and $a$ is the shrinkage-dependent contact radius $\sqrt{\delta R/2}$, consistent with our observations of thin capillary menisci between beads \cite{supp,doi:10.1021/acs.langmuir.5b04012,doi:10.1021/la401115j,doi:10.1021/la702633t,Moller:2007bp,doi:10.1021/la000657y,Cho2019cracks1}. We neglect inter-bead friction since this has been measured to be negligible \cite{doi:10.1063/1.4983047}. The balance between the contact and capillary forces then determines $\delta = 8\pi \gamma / \left( 3K \right)$ (schematized in Fig.~\ref{fig:fig1}b) as previously validated \cite{supp, Moller:2007bp,Fogden:1990ej,Butt:2010gs,Herminghaus:2005ei,PhysRevE.84.011302,PhysRevE.82.041403,Cho2019cracks1}. As intuitively expected, the bead compression depends inversely with the stiffness. To incorporate the influence of bead size, we further express this compression in nondimensional form at the initial hydrated state (Fig.~S3): 
\begin{equation}
	\hat{\delta}_\text{wet} \equiv \frac{\delta_\text{wet}}{R_\text{wet}} = \frac{8\pi\gamma}{3K_\text{wet} R_\text{wet}} \text{.}
	\label{eqn:deltahatwet}
\end{equation}
This parameter thus describes the combined influence of bead compression and bead size on cracking; in each DEM simulation, we specify the value of $\hat{\delta}_\text{wet}$ to be tested.

Cracking also depends on the competition between capillary cohesion holding the packing together and frictional forces at the substrata immobilizing the grains. Our simulations treat friction on a single-bead basis by first determining the net capillary and contact forces on a bead. If this net force exceeds a constant static friction threshold $F_\text{fr}$---which we specify in each DEM simulation---the bead moves; otherwise, it does not. 

Conversely, $F_\text{fr}$ is similar for all the experiments, and the relative importance of friction can instead be tuned by changing the packing height. This dependence can be intuitively understood by considering the stress at the boundary of a cluster just before cracking,
\begin{equation}
  \sigma_\text{c} = \frac{\text{friction}\times \text{cluster area}}{\text{cluster surface area}} \sim \frac{N F_\text{fr}}{N^{1/2}H R^2} \text{,}
    \label{eqn:sigma}
\end{equation}
which quantifies the diminishing role that friction plays as packing height increases; here, $H$ is the number of bead monolayers making up the packing height. Thus, to compare results from the simulations and the experiments, we combine the effects of friction and height into a single nondimensional parameter, the \emph{effective height}:
\begin{equation}
	\mathcal{H} \equiv H \frac{2\pi\gamma R_\text{wet} }{ F_\text{fr} }\text{.}
	\label{eqn:effheight}
\end{equation}
The simulations thus represent a 2D cross-section of the 3D experimental packing.



The DEM simulations solve for bead sizes, bead positions, and capillary-bridge bonds by explicitly treating bead-scale contact, capillary, and friction forces during drying \cite{Cho2019cracks1}. To explicitly treat the evolution of bead-scale parameters with shrinkage, we use moduli that change as the beads shrink: $K=K_\text{wet}\left(R_\text{wet}/R\right)^{3m}$ with $m = 9/4$ as verified by others \cite{deGennes:1979uw,ZRINYI19871139,vanderSman:2015ef,doi:10.1021/ma011408z}. Moreover, to match the experiments, we ensure that the simulated bead shrinkage is uniform during drying; thus, our simulations do not probe dynamic effects arising from non-uniform water content \cite{supp, Cho2019cracks1}. We implement a gradient-descent scheme to find the mechanical minimum for the beads as the packing dries, and use a hybrid event-detection scheme to break bonds when they become overstretched. Both initially ordered and disordered structures yield similar results (closed and open symbols in Fig.~\ref{fig:demresults}d), indicating that our results can be generalized.

To characterize the crack patterns in the DEM simulations, we count the number of missing capillary bonds per monolayer for the final cracked state, $u$. This quantity yields a direct measure of total crack length, including the packing boundary as in the experimental analysis. Again assuming uniform clusters then yields the characteristic cluster area, $N \sim M^2/u^2$, where $M$ is the total number of beads in the 2D packing cross-section \cite{supp}. 

We use the DEM simulations to directly probe the influence of varying $\Phi$ and $\hat{\delta}_\text{wet}$---which quantify grain shrinkability, stiffness, and size---on cracking. We hold the nondimensional friction $F_\text{fr}/\left(2\pi\gamma R_\text{wet}\right)$ fixed to mimic the experiments. Remarkably, we find similar crack patterns in the simulations (Fig.~\ref{fig:demresults}a--c) and the experiments (Fig.~\ref{fig:fig1}d--f). In particular, the total crack length $u$ increases, and the equivalent cluster area $N$ decreases, with increasing $\Phi$; compare Fig.~\ref{fig:demresults}a,b. The crack length also increases, and the cluster area decreases, with decreasing $\hat{\delta}_\text{wet}$ and increasing $R_\text{wet}$; compare Fig.~\ref{fig:demresults}b,c. Our simulations thus reproduce the experimental observations, indicating that they successfully capture the key underlying physics. 

The DEM simulations also provide an opportunity to test the relative importance of substrate friction on cracking by tuning $F_\text{fr}$, and thus, $\mathcal{H}$. Previous work has shown that the equivalent cluster area scales with packing height as a power law: $N\propto \mathcal{H}^{4/3}$ \cite{Flores:2017js,PhysRevE.99.012802}. We observe a similar scaling in our simulations for all packing geometries, bead shrinkabilities, stiffnesses, and sizes tested (Fig.~\ref{fig:demresults}d), suggesting that a similar balance of forces mediates cracking. However, our results demonstrate that the prefactor to this scaling also depends on the previously-overlooked parameter $\Phi$, as shown in Fig.~\ref{fig:demresults}e. Specifically, we find that cluster area can vary over an order of magnitude with varying $\Phi$, even with all other parameters unchanged---an effect that is missed in previous work. Clearly, the current understanding of cracking is incomplete.

To more completely describe cracking, we explicitly incorporate $\Phi$ into classical crack theory. Following previous work \mbox{\cite{Flores:2017js,PhysRevE.99.012802}}, we consider the balance between the surface energy required to create a crack and the strain energy within a cluster for 2D clusters that are extruded along the packing height \cite{supp}:
\begin{equation}
  \underbrace{\frac{2\pi \gamma a \delta}{R^2}}_{\substack{\text{crack surface energy} \\ \text{per unit area}}} \underbrace{N^{1/2} R^2 H}_{\substack{\text{cluster} \\ \text{surface area}}} ~\sim \underbrace{\frac{\sigma_\text{c}^2}{2 \mathcal{E}}}_{\substack{\text{strain energy} \\ \text{per unit volume}}} \underbrace{N R^3 H}_{\substack{\text{cluster} \\ \text{volume}}} \text{.}
  \label{eqn:griffithenergybalance}
\end{equation}
Here, $a=\sqrt{\delta R/2}$ is the contact radius between equally sized beads, $\mathcal{E}=\left(3/16\right)\sqrt{3 K \gamma / \left(\pi R\right)}$ is the effective Young's modulus of a pendular packing \mbox{\cite{Cho2019cracks1,Moller:2007bp}}, and $\sigma_\text{c}$ is the stress resulting from friction with the substrata at the point of cracking (Equation~{\ref{eqn:sigma}}). Importantly, both the surface energy and the strain energy depend on grain shrinkability $\Phi$; these dependencies are not incorporated in previous descriptions of cracking for continuum solids, because they arise from grain-scale processes. We assume that at the point of cracking, the granular properties are determined by the final dehydrated state: $R\sim R_\text{dry}$, $K\sim K_\text{dry}$, $a\sim a_\text{dry}=\sqrt{\delta_\text{dry} R_\text{wet}/\left(2\Phi^{1/3}\right)}$ and $\delta \sim \delta_\text{dry}=R_\text{wet} \hat{\delta}_\text{wet} \Phi^{-m}$, using Eq.~{\ref{eqn:deltahatwet}}. Thus, the Griffith energy balance (Eq.~\ref{eqn:griffithenergybalance}) can be expressed completely in terms of $N$ and $\mathcal{H}$, as well as the grain-scale parameters $\Phi$ and $\hat{\delta}_\text{wet}$. Solving for $N$ and applying Eqs.~\ref{eqn:deltahatwet}, \ref{eqn:sigma}, and \ref{eqn:effheight} then yields a general scaling law for cluster size:
\begin{equation}
  N \sim  \hat{\delta}_\text{wet}^{2/3} \Phi^{-(2m/3+2/9)} \mathcal{H}^{4/3} \text{,}
  \label{eqn:scalinglaw}
\end{equation}
where the exponent $m=9/4$ for our hydrogels, but can be modified for other shrinkable materials. This scaling yields the $N \propto \mathcal{H}^{4/3}$ dependence found in other studies \cite{Flores:2017js,PhysRevE.99.012802}, but also explicitly describes the dependence of cracking on bead shrinkability. Specifically, we expect that the propensity for cracking increases, and thus the equivalent cluster area decreases, as bead shrinkability $\Phi$ increases: the capillary bridges between beads are more likely to stretch and break, resulting in cracking. 


\begin{figure}[htb]
\centering
  \includegraphics[width=8.65cm]{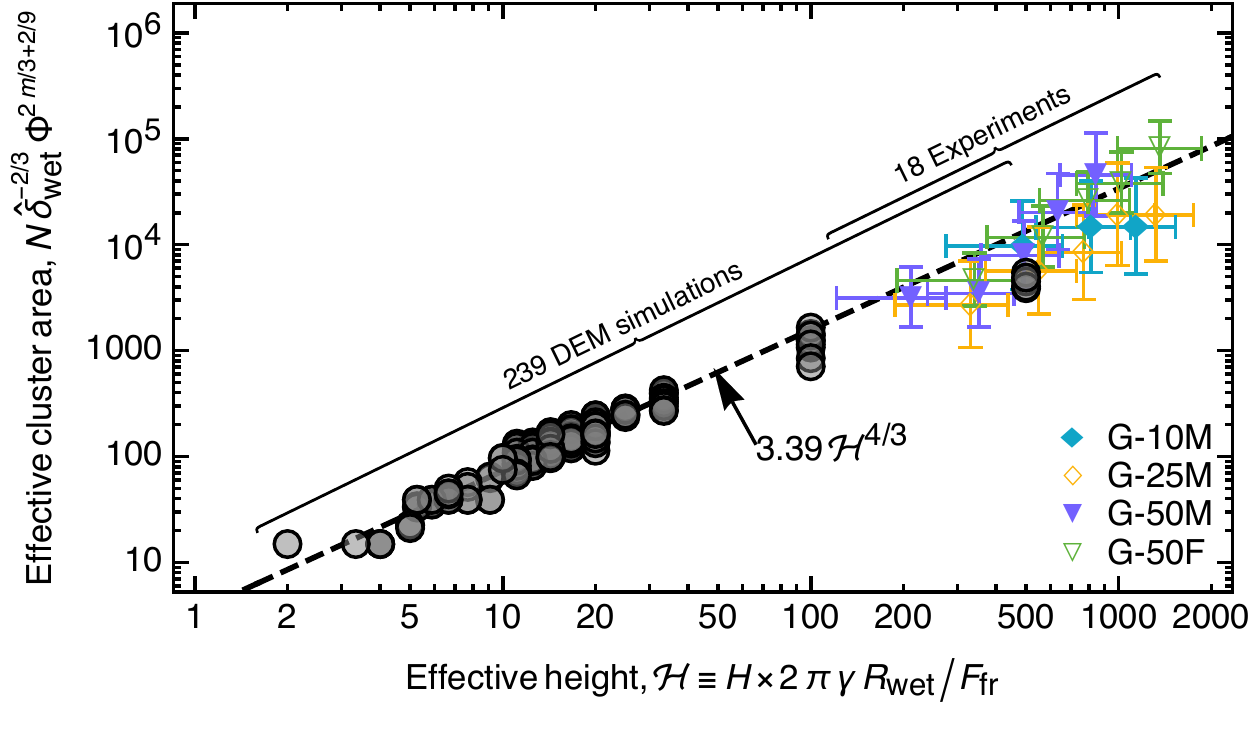}
  \caption{DEM simulation and experimental results all collapse on our scaling law (Eq.~{\ref{eqn:scalinglaw}}). We perform a regression on the simulation data to obtain the dashed line, with a coefficient of determination = 0.98. For the experimental data, we fix $F_\text{fr}=0.02\times2\pi\gamma R_\text{wet}$, in independent agreement with our measurement of friction. Error bars reflect the combined uncertainties in crack length, bead size, and the packing fraction $\eta$. Different symbols show different classes of Sephadex beads with different shrinkability, compression, and bead size. We assume $m=9/4$.}
  \label{fig:finalcollapse}
\end{figure}

Our DEM simulations---which explicitly incorporate the nonlinear bead interactions and shrinkage-dependent bead properties---independently probe $\mathcal{H}$, $\Phi$, and $\hat{\delta}_\text{wet}$ varying over one to three orders of magnitude (Fig.~{\ref{fig:demresults}}d--f). These results enable us to directly test Eq.~{\ref{eqn:scalinglaw}}. Despite our simplifying assumptions of linear elasticity and quasistatic cracking, we find that this scaling law collapses all of the results onto a single curve, as shown by the circles in Fig.~{\ref{fig:finalcollapse}}. Moreover, performing a regression---fitting only a constant of proportionality---on all of the results yields the relation $N = 3.39~\hat{\delta}_\text{wet}^{2/3} \Phi^{-(2m/3+2/9)} \mathcal{H}^{4/3}$, in excellent agreement with the scaling law. As a final verification of our scaling law, we perform experiments on hydrogel bead packings of systematically varying $\Phi$, $\hat{\delta}_\text{wet}$, and height, measuring $N$ in each case. We again find that Eq.~\ref{eqn:scalinglaw} collapses all the results onto the same curve as the DEM results, as shown by the symbols in Fig.~\ref{fig:finalcollapse}. This agreement between the experiments, DEM simulations, and theoretical prediction provides strong evidence of the validity of our new description of cracking.


Our work sheds light on how cracking depends on the interplay between grain shrinkage, stiffness, and size, as well as capillary cohesion and substrate friction. By explicitly incorporating grain shrinkage into classical crack theory, our scaling prediction provides a way to predict crack patterning in diverse shrinkable, granular packings---including clays, soils, coatings, biological tissues, and foods. We anticipate that our work will also guide controlled cracking for applications such as micro/nano-patterning \cite{Kim:2016bh} and transport in fuel cells \cite{Kim:2016gr}.

It is a pleasure to acknowledge M. P. Howard for helpful discussions that guided the development of the DEM approach, N. B. Lu for initial experimental observations that motivated this work, and W. B. Russel and G. W. Scherer for stimulating discussions at the inception of this work. We also acknowledge the Princeton Institute for Computational Science and Engineering for computer cluster access. This work was supported by start-up funds from Princeton University, the Alfred Rheinstein Faculty Award, the Grand Challenges Initiative of the Princeton Environmental Institute, and in part by funding from the Princeton Center for Complex Materials, a Materials Research Science and Engineering Center supported by NSF grant DMR-1420541.

\end{document}